\documentclass[10pt,aps,preprintnumbers,prd,noshowpacs,nofootinbib,noshowkeys,floatfix,superscriptaddress]{revtex4}
\usepackage[dvips]{graphics,graphicx}
\usepackage[colorlinks=true,linktocpage=true,linkcolor=blue,citecolor=blue]{hyperref}
\usepackage[usenames,dvipsnames]{color}
\usepackage{amsmath, amssymb,oldgerm}
\usepackage{multirow}
\usepackage{xcolor}
\usepackage[normalem]{ulem}  
\usepackage{braket}
\usepackage{slashed}
\usepackage[mathscr]{eucal}
\usepackage[dvips]{graphics,graphicx}
 \usepackage{caption}
\usepackage{subcaption}

\newcommand{\n}{\nonumber}

\newcommand{\mC}{\mathfrak{C}}

\newcommand{\ms}{\mathfrak{s}}

\newcommand{\beq}{\begin{equation}}
\newcommand{\eeq}{\end{equation}}

\newcommand{\eqs}{Eqs.~}
\newcommand{\eq}{Eq.~}

\newcommand{\taum}{\tau}
\newcommand{\proj}{\Delta}

\newcommand{\tc}{\mathfrak{K}}

\newcommand{\lmur}{{\langle\mu\rangle}}

\begin{document}

\title{Linearly stable and causal relativistic first-order spin hydrodynamics}

\author{Nora Weickgenannt}
\email{nora.weickgenannt@ipht.fr}

\affiliation{Institut de Physique Th\'eorique, Universit\'e Paris Saclay, CEA, CNRS, F-91191 Gif-sur-Yvette, France}

\begin{abstract}
 We derive equations of motion for dissipative spin hydrodynamics from kinetic theory up to first order in a gradient expansion. Choosing a specific form of the matching conditions, relating the change in the spin potential to the spin diffusion and spin energy, we then show that the equations of motion, linearized around homogeneous global equilibrium, are causal and stable in any Lorentz frame, if certain sufficient conditions on the transport coefficients are fulfilled.
\end{abstract}

\maketitle

\section{Introduction}

Relativistic spin hydrodynamics~\cite{Florkowski:2017ruc,Florkowski:2017dyn,Montenegro:2017rbu,Florkowski:2018myy,Florkowski:2018fap,Montenegro:2018bcf,Hattori:2019lfp,Bhadury:2020puc,Bhadury:2020cop,Singh:2020rht,Gallegos:2020otk,Garbiso:2020puw,Montenegro:2020paq,Speranza:2020ilk,Gallegos:2021bzp,Fukushima:2020ucl,Li:2020eon,Bhadury:2021oat,Peng:2021ago,Wang:2021ngp,Wang:2021wqq,Hu:2021pwh,Hu:2022lpi,Hongo:2021ona,Singh:2022ltu,Daher:2022xon,Weickgenannt:2022zxs,Ambrus:2022yzz,Weickgenannt:2022jes,Gallegos:2022jow,Bhadury:2022ulr,Cao:2022aku,Weickgenannt:2022qvh,Biswas:2023qsw} is a recently developed theory, mainly motivated by the observation of polarization phenomena in heavy-ion collisions~\cite{Liang:2004ph,Voloshin:2004ha,Betz:2007kg,Becattini:2007sr,STAR:2017ckg,Adam:2018ivw,ALICE:2019aid,Mohanty:2021vbt}. In the framework of spin hydrodynamics, the spin tensor is treated as an independent dynamical variable, with its equations of motion following from the conservation of total angular momentum~\cite{Florkowski:2017ruc,Florkowski:2017dyn}. If the system is in local equilibrium, the spin tensor may be uniquely~\footnote{The form of the spin tensor depends on the choice of pseudo-gauge~\cite{Hehl:1976vr}. Once a pseudo-gauge is chosen, the microscopic definition of the spin tensor in this pseudo-gauge is a well-defined function of the spin potential in local equilibrium.} expressed as a function of the so-called spin potential, which serves as the thermodynamic potential for the conservation of total angular momentum. The six equations of motion obtained from the macroscopic conservation law for the angular-momentum tensor then provide a closed system of equations of motion, determining the spin potential, and hence the spin tensor. On the other hand, as soon as dissipative effects are included in the theory, all components of the spin tensor are independent, and one requires additional equations in order to determine the unknown components.

There are two common strategies to deal with such an issue in a conventional hydrodynamic theory. In so-called first-order hydrodynamics, one expresses all dissipative components of the conserved currents, i.e., of the charge current and the energy-momentum tensor, by space-time derivatives of the thermodynamic potentials, given by the inverse temperature $\beta$, the chemical potential over temperature $\alpha$, and the fluid velocity $u^\mu$, up to first order in a gradient expansion. Then, one obtains closed equations of motion for the five independent quantities $\alpha$, $\beta$, and $u^\mu$, which constitute the only dynamical variables of the theory. A well known example for such a theory is Navier-Stokes hydrodynamics~\cite{eckart1940thermodynamics}. Unfortunately, relativistic Navier-Stokes equations of motion feature unphysical properties, such as instability of global equilibrium and violation of causality~\cite{hiscock1985generic}. Recently, it has been found that such unphysical behavior is not a general feature of first-order theories, but instead related to the choice of matching conditions, which define the thermodynamic potentials in a nonequilibrium situation. In Bemfica-Disconzi-Noronha-Kovtun (BDNK) hydrodynamics, causality and stability can be guaranteed, given that certain conditions on the matching coefficients are fulfilled~\cite{Bemfica:2017wps,Bemfica:2019knx,Kovtun:2019hdm,Bemfica:2020zjp,Hoult:2020eho}. The discovery of this new class of hydrodynamic theories lead to a series of recent investigations~\cite{Das:2020grz,Speranza:2021bxf,Hoult:2021gnb,Biswas:2022cla,Rocha:2022ind,Bemfica:2022dnk,Bemfica:2023res,Rocha:2023hts}.

The other type of causal and stable hydrodynamic theories is known as second-order hydrodynamics.~\footnote{We follow the literature by using the terms "first-order" and "second-order" hydrodynamics, although second-order hydrodynamics is not obtained by adding second-order terms to first-order hydrodynamics. The crucial difference is the number of dynamical variables, which are five for first-order, but 14 for second-order hydrodynamics.} In this approach, all 14 components of the conserved currents are treated dynamically and follow their own equations of motion~\cite{Israel:1979wp,Denicol:2012cn}. Second-order relativistic hydrodynamics is established as an effective theory to describe the dynamics in heavy-ion collision~\cite{Heinz:2013th,Florkowski:2017olj}, in particular, the equations of motion for the 14 dynamical moments can describe the evolution of the system already in a regime far from local equilibrium due to the approximately boost-invariant expansion in heavy-ion collisions~\cite{Heller:2015dha,Blaizot:2017ucy}. On the other hand, for systems sufficiently close to local equilibrium, BDNK hydrodynamics is expected to yield a good description of the dynamics, while being simpler than second-order hydrodynamics. Another advantage of BDNK hydrodynamics is that causality and stability of the equations of motion can be easier controlled by adjusting the matching conditions. Therefore, when deriving equations of motion for spin hydrodynamics, it is desirable to include spin degrees of freedom in both formulations, and decide in a specific application whether first- or second-order spin hydrodynamics is more suitable to describe the respective situation.

First-order spin hydrodynamics with Navier-Stokes type matching conditions has been obtained in Refs.~\cite{Hattori:2019lfp,Hu:2021pwh} Recently, it has been shown that this theory suffers from instability and acausality~\cite{Sarwar:2022yzs,Daher:2022wzf,Biswas:2022bht,Xie:2023gbo}, similar to conventional Navier-Stokes theory. Furthermore, second-order spin hydrodynamics has been derived from spin kinetic theory~\cite{Weickgenannt:2019dks,Weickgenannt:2020aaf,Weickgenannt:2021cuo,Wagner:2022amr} [see also Refs.\cite{Yang:2020hri,Wang:2020pej,Sheng:2021kfc,Sheng:2022ssd,Wagner:2022gza,Wagner:2023cct} for related work] in Refs.~\cite{Weickgenannt:2022zxs,Weickgenannt:2022qvh} and from an entropy-current analysis in Ref.~\cite{Biswas:2023qsw}. The equations of motion in these theories are expected to be causal and stable for a certain range of transport coefficients, which is still to be investigated, see Refs.~\cite{Sarwar:2022yzs,Xie:2023gbo} for first studies in this direction. On the other hand, up to now a causal and stable formulation of relativistic first-order spin hydrodynamics does not exist. The goal of this paper is to fill this gap, and provide BDNK-type equations of motion for spin hydrodynamics together with conditions on the transport coefficients, which, if fulfilled, guarantee causality and stability of the theory, at least in the linear regime around global equilibrium.

In this paper, we derive equations of motion for first-order spin hydrodynamics from kinetic theory, using the methods established in Refs.~\cite{Weickgenannt:2022zxs,Weickgenannt:2022qvh}. Assuming the collision term to be local, we obtain the equations of motion for the spin potential from the conservation of the spin tensor. Non-equilibrium contributions to the latter, which appear in the conservation equation, are expressed in terms of gradients of the thermodynamic potentials by making use of the Boltzmann equation up to first order. There are six dissipative components of the spin tensor for which this strategy does not apply, since the projections of the Boltzmann equation onto the relevant subspace in momentum space are redundant with the already used conservation law for the spin tensor. For this reason, one requires matching conditions. We choose a matching which relates the respective components of the spin tensor, corresponding to spin energy and spin diffusion, to time-like derivatives of the spin potential. While this is not the most general form of matching, it is sufficient to render the equations of motion for the spin potential causal and stable. In order to demonstrate this, we linearize the resulting equations of motion around homogeneous, i.e., non-rotating and unpolarized, global equilibrium. Expanding the perturbations of the spin potential in linear modes, we obtain dispersion relations for the solutions. We then derive conditions on the transport coefficients, which, if fulfilled, guarantee the causality and stability of the theory in the linear regime in any Lorentz frame.

This paper is organized as follows. In Section \ref{kinsec}, we derive first-order equations of motion for the spin potential, and linearize them around equilibrium. In Section \ref{modesec}, we perform the linear mode expansion, which splits up into longitudinal and transverse spin modes. The stability and causality conditions obtained from the dispersion relations of the longitudinal spin modes are discussed in Section \ref{longsec}, while the same is done for the transverse spin modes in Section \ref{transsec}. Finally, a summary of the stability and causality conditions and conclusions are provided in Section \ref{concsec}.

We use the following notation and conventions, $a\cdot b=a^\mu b_\mu$,
$a_{[\mu}b_{\nu]}\equiv a_\mu b_\nu-a_\nu b_\mu$, $a_{(\mu}b_{\nu)}\equiv a_\mu b_\nu
+a_\nu b_\mu$, $g_{\mu \nu} = \mathrm{diag}(+,-,-,-)$,
$\epsilon^{0123} = - \epsilon_{0123} =\epsilon^{123}= 1$, and repeated indices are summed over. 
The dual of any rank-2 tensor $A^{\mu\nu}$ is defined as 
$\tilde{A}^{\mu\nu}\equiv \epsilon^{\mu\nu\alpha\beta}A_{\alpha\beta}$. We do not distinguish between upper and lower spatial indices of three vectors.

\section{First-order spin hydrodynamics from kinetic theory}
\label{kinsec}

Our starting point to derive equations of motion for dissipative spin hydrodynamics is the Boltzmann equation~\cite{Weickgenannt:2020aaf}
\begin{equation}
    p\cdot\partial f(x,p,\ms)=\mC[f]\; ,
    \label{boltz}
\end{equation}
where $f(x,p,\ms)$ is the distribution function in extended phase space and $\mC[f]$ is the collision term. The conserved quantities in spin hydrodynamics, which we aim to calculate from the equations of motion to be derived in the following, are
the charge current $N^{\mu}$, the Hilgevoord-Wouthuysen (HW) energy-momentum tensor $T^{\mu\nu}$, and the HW spin tensor $S^{\lambda,\mu\nu}$, given by~\cite{Speranza:2020ilk}
\begin{align}
N^\mu &= \left\langle p^\mu \right\rangle\;,\n\\
T^{\mu\nu}&= \left\langle p^\mu p^\nu \right\rangle\;, \n\\
S^{\lambda,\mu\nu}&= \frac12\left\langle p^\lambda  \Sigma_{\ms}^{\mu\nu}\right\rangle 
-\frac{\hbar}{4m^2}\partial^{[\nu} \left\langle p^{\mu]} p^\lambda \right\rangle\;.  \label{spintens}
\end{align}
Here we introduced
$\langle \cdots \rangle\equiv \int d\Gamma\, (\cdots) f(x,p,\ms)\;,$
with
  $d\Gamma  \equiv dP\, d S(p)$
denoting integration over the extended phase space, where the on-shell momentum measure is given by
    $dP\equiv (d^3p/2p^0) $
and the spin measure reads
\begin{equation} \label{dSp}
dS(p) \equiv \frac{m}{\sqrt{3}\pi}  d^4\ms\,  \delta(\ms\cdot\ms+3)\delta(p\cdot \ms)\;.
\end{equation}
We also defined the dipole-moment tensor
\begin{equation}
\Sigma_\ms^{\mu\nu}\equiv -\frac{1}{m} \epsilon^{\mu\nu\alpha\beta} p_\alpha \ms_\beta\;. 
\end{equation}
In general, the HW energy-momentum tensor may have a nonzero antisymmetric part, originating from a nonlocal collision term, and leading to a nonconserved spin tensor~\cite{Weickgenannt:2020aaf}. However, in this work we will consider a local collision term, such that the spin tensor is conserved,
\begin{equation}
    \partial_\lambda S^{\lambda,\mu\nu}= 0\; . \label{partials}
\end{equation}
Also note that the second term in the last equation in \eqref{spintens} is conserved separately due to the conservation of the energy-momentum tensor,
\begin{equation}
    \partial_\lambda T^{\mu\lambda}=0\; . \label{partialt}
\end{equation}
Therefore, the conserved current related to the conservation of total angular momentum is in our case given by the first term in the expression for the spin tensor,
\begin{equation}
S_\Sigma^{\lambda,\mu\nu}\equiv\frac12\left\langle p^\lambda  \Sigma_{\ms}^{\mu\nu}\right\rangle \; .
\label{spintenssigma}
\end{equation}
In the following, we will show how to obtain this quantity within a gradient expansion around local equilibrium up to first order. 
To this end, we expand the distribution function $f(x,p,\ms)$ in dissipative spin moments  $\tau_n^{\lmur,\mu_1\cdots\mu_\ell}$~\cite{Weickgenannt:2022zxs},
\begin{align}
& f(x,p,\ms) =f_\text{eq}(x,p,\ms)-f^{(0)}_\text{eq}(x,p)
\sum_{l=0}^\infty \sum_{n\in \mathbb{S}_l}\mathcal{H}_{pn}^{(l)}\left(g_{\mu\nu}-\frac{p_{\langle\mu\rangle}}{E_p} u_\nu \right)
 \ms^\nu \tau_n^{\lmur,\mu_1\cdots\mu_l}p_{\langle\mu_1}\cdots p_{\mu_l\rangle}\;, \label{distrnoneqfin}
 \end{align} 
where $E_p\equiv p\cdot u$, $p^{\langle\mu\rangle}\equiv \proj^{\mu\nu}p_\nu$, $p_{\langle\mu_1}\cdots p_{\mu_l\rangle}$ are irreducible tensors in momentum space, $\mathbb{S}_l$ is the set of considered spin moments, and the functions $\mathcal{H}_{pn}^{(l)}$ are defined in \eq\eqref{Hcoeffdef}.
The local-equilibrium distribution function up to first-order in $\hbar$ is given by~\cite{Weickgenannt:2020aaf}
\begin{equation}
f_\text{eq}(x,p,\ms)=\frac{1}{(2\pi\hbar)^3} e^{-\beta\cdot p+\alpha}\left(1+\frac\hbar4\Omega_{\mu\nu}\Sigma_\ms^{\mu\nu} \right)\; ,
\end{equation}
and $f^{(0)}_\text{eq}(x,p)$ denotes the local-equilibrium distribution function at zeroth order in $\hbar$. Furthermore, $\beta^\mu\equiv\beta u^\mu$, $u^\mu$ is the fluid velocity, $\beta$ is the inverse temperature, $\alpha$ is the chemical potential over temperature, and the antisymmetric tensor $\Omega_{\mu\nu}$ is the spin potential.

In Refs.~\cite{Weickgenannt:2022zxs,Weickgenannt:2022qvh} we derived equations of motion for second-order dissipative spin hydrodynamics from kinetic theory using the method of moments. In that approach, the 24 spin moments constituting the components of the HW spin tensor \eqref{spintens} are treated dynamically, following equations of motion derived from the Boltzmann equation. On the other hand, in this work, we will follow a different strategy and express the spin moments as a function of the spin potential up to first order in gradients. Therefore, only the six independent components of the spin potential constitute the dynamical variables of the theory, for which we will derive equations of motion in the following. The spin tensor can then be expressed in terms of the spin potential up to first order in gradients. An infinite set of coupled equations of motion for both the spin potential and the spin moments has been obtained in Ref.~\cite{Weickgenannt:2022zxs} from the Boltzmann equation \eqref{boltz} without further approximations and will be used as the starting point for the derivation of first-order spin hydrodynamics in this work. The only difference concerning these equations lies in the matching conditions, which are essential to obtain causal and stable first-order hydrodynamics~\cite{Bemfica:2017wps,Bemfica:2019knx,Kovtun:2019hdm,Bemfica:2020zjp,Hoult:2020eho}. In second-order spin hydrodynamics, it is convenient to choose a Landau-type matching condition of the form
\begin{equation}
    u_\lambda J^{\lambda,\mu\nu}=u_\lambda J^{\lambda,\mu\nu}_\text{eq}\; , \label{landauJ}
\end{equation}
where 
\begin{equation}
    J^{\lambda,\mu\nu}\equiv x^\mu T^{\lambda\nu}-x^\nu T^{\lambda\mu}+\hbar S^{\lambda,\mu\nu}
\end{equation}
is the total angular-momentum tensor.
This has been done in Refs.~\cite{Weickgenannt:2022zxs,Weickgenannt:2022qvh}. For a local collision term, as considered in this work, and in combination with the standard Landau matching condition~\footnote{The matching condition for the energy-momentum tensor is also called hydrodynamic frame, e.g., \eq\eqref{landauT} corresponds to the Landau frame.}
\begin{equation}
    u_\lambda T^{\lambda\mu}=u_\lambda T^{\lambda\mu}_\text{eq}
    \label{landauT}
\end{equation}
the condition \eqref{landauJ} is equivalent to
\begin{equation}
    u_\lambda S_\Sigma^{\lambda,\mu\nu}=u_\lambda S^{\lambda,\mu\nu}_{\Sigma,\text{eq}}\; . \label{landauS}
\end{equation}
On the other hand, in order to obtain causal and stable first-order hydrodynamics, it is essential to choose matching conditions different from \eqs\eqref{landauT}~\cite{Bemfica:2017wps,Bemfica:2019knx,Kovtun:2019hdm,Bemfica:2020zjp,Hoult:2020eho} and \eqref{landauS}, as we will see in the following [see also recent discussions in Refs.~\cite{Sarwar:2022yzs,Daher:2022wzf,Biswas:2022bht,Xie:2023gbo} for instability of spin hydrodynamics with Landau-type matching]. We will therefore relax conditions \eqref{landauT} and \eqref{landauS} and add the respective terms to the equations of motion obtained in Ref.~\cite{Weickgenannt:2022zxs}

It is convenient to define the following components of the spin potential,
\begin{equation}
    \omega_0^\alpha\equiv \frac12\epsilon^{\alpha\beta\mu\nu} u_\beta \Omega_{\mu\nu}
\end{equation}
and
\begin{equation}
    \kappa_0^\mu\equiv -\Omega^{\mu\nu}u_\nu\; .
\end{equation}
The equations of motion for $\omega_0^\alpha$ and $\kappa_0^\mu$ are then derived by inserting \eq\eqref{distrnoneqfin} into the conservation law for the spin tensor \eqref{partials} and performing the momentum integration with the orthogonality relation \eqref{orthrel_poly}, c.f.\ Ref.~\cite{Weickgenannt:2022zxs},
\begin{align}
\frac{\hbar}{m^2}\dot{\omega}_0^{\langle\alpha\rangle}=& -\frac{2}{I_{30}- I_{31}} 
\left\{  \left[ \frac{\hbar}{2m^2} ( 2\theta I_{30}-I_{40}\dot{\beta}_0+I_{30}\dot{\alpha}_0)
+\frac{\hbar}{2m^2}( I_{41}\dot{\beta}_0-I_{31}\dot{\alpha}_0)-\frac16\frac{\hbar}{m^2}I_{31}\theta\right]
\omega_0^\alpha\right.\n\\
& +u_\lambda\nabla^\alpha\frac{1}{3m}\left(m^2 \tau_{0}^\lambda-\taum_2^\lambda\right)
-u_\lambda \frac{1}{3m}\left(m^2 \tau_{0}^\lambda-\taum_2^\lambda\right)\dot{u}^\alpha
+\frac{1}{3m} \theta \left(m^2 \tau_{0}^{\langle\alpha\rangle}-\taum_2^{\langle\alpha\rangle}\right) \n\\
&\left.-\frac12\frac{\hbar}{m^2} \epsilon^{\langle\alpha\rangle\lambda\mu\nu}\kappa_{0\nu}
\left( -I_{41}u_\mu\nabla_\lambda\beta_0+I_{31}u_\mu\nabla_\lambda\alpha_0
 +3I_{31} u_\mu \dot{u}_\lambda \right)-\frac12\frac{\hbar}{m^2}\epsilon^{\langle\alpha\rangle\lambda\mu\nu}I_{31} 
u_\mu\nabla_\lambda\kappa_{0\nu}\right.\n\\
&\left.-\frac{\hbar}{m^2} I_{31} (\sigma^{\alpha\lambda}+\omega^{\alpha\lambda})\omega_{0\lambda}-\frac{1}{2m} \proj^\alpha_\beta \nabla_\lambda 
\taum_1^{\rho,(\lambda}\proj^{\beta)}_\rho +\frac{1}{2m} \taum_1^{\langle\alpha\rangle,\nu}\dot{u}_\nu  +\frac1m u_\beta \proj^\alpha_\rho\nabla_\lambda 
\taum_0^{[\beta,\rho]\lambda} \right.\n\\
&\left. -\frac1m u_\beta \taum_0^{\beta,\alpha\lambda}\dot{u}_\lambda
 -\frac1m\theta\left(\tau_2^{\langle\alpha\rangle}-u_\beta\tau_1^{\beta,\alpha}\right)-\frac1mu_\beta\proj^\alpha_\lambda\frac{d}{d\tau}\left(\tau_2^{[\lambda}u^{\beta]}-\tau_1^{[\beta,\lambda]}\right)\right\}\;, \label{omegadot2}
\end{align}
and
\begin{align}
\frac{\hbar}{m^2}\dot{\kappa}^{\langle\mu\rangle}_0=&  -\frac{1}{I_{31}} \bigg\{  
\frac{\hbar}{2m^2}I_{30} \epsilon^{\mu\nu\alpha\beta} \dot{u}_\alpha \omega_{0\beta} u_\nu+\frac{\hbar}{2m^2}\epsilon^{\alpha\mu\nu\beta} u_\nu 
\left[-I_{31} \nabla_\alpha \omega_{0\beta}+(I_{41}\nabla_\alpha\beta_0
-I_{31}\nabla_\alpha \alpha_0)\omega_{0\beta}\right]\n\\
&-{\frac{1}{3m}}\epsilon^{\alpha\mu\nu\beta}u_\nu\nabla_\alpha \left(m^2\taum_{0\beta}
-\taum_{2\beta}\right)+\frac{1}{3m}\epsilon^{\alpha\mu\nu\beta}u_\nu\dot{u}_\alpha 
\left(m^2\taum_{0\beta}-\taum_{2\beta}\right)-\frac{\hbar}{2m^2} I_{31}(\sigma^{\mu\nu}+\omega^{\mu\nu})\kappa_{0\nu}
\n\\
&+\frac{1}{2m} \epsilon^{\mu\nu\alpha\beta} u_\alpha \taum_{1(\beta,\lambda)}
\left(\sigma^\lambda_{\ \nu}+\omega^\lambda_{\ \nu}\right)
+\frac{\hbar}{m^2}\left( \frac43 I_{31} \theta-I_{41}\dot{\beta}_0+I_{31}\dot{\alpha}_0 \right)
\kappa_0^\mu\n\\
&-\frac{1}{m} \epsilon^{\mu\nu\alpha\beta} u_\nu \left(\nabla^\lambda \taum_{0\beta,\alpha\lambda}
-\dot{u}^\lambda \taum_{0\beta,\alpha\lambda} \right)-\frac1m\theta\epsilon^{\mu\nu\alpha\beta}u_\nu  \tau_{1\beta,\alpha}-\frac1m\epsilon^{\mu\nu\alpha\beta}u_\nu \frac{d}{d\tau} (u_\alpha \tau_{2\beta}+\tau_{1\beta,\alpha}) \bigg\}\; .
\label{kappadot2}
\end{align}
Here we defined $\nabla^\mu\equiv \proj^\mu_\nu \partial^\nu$ and 
$\dot{A}\equiv u\cdot \partial A \equiv dA/d\tau$, as well as the expansion scalar 
$\theta\equiv \nabla\cdot u$, the shear tensor 
$\sigma^{\mu\nu}\equiv \nabla^{\langle\mu} u^{\nu\rangle}\equiv [(1/2)\proj^{(\mu}_\alpha\proj^{\nu)}_\beta-(1/3)\proj^{\mu\nu}\proj_{\alpha\beta}]\nabla^\alpha u^\beta $, the fluid vorticity 
$\omega^{\mu\nu}\equiv (1/2)\nabla^{[\mu} u^{\nu]}$, and the thermodynamic integrals $I_{nq}$ given in \eq\eqref{thermdynint}.
Note that \eqs\eqref{omegadot2} and \eqref{kappadot2} differ from the corresponding equations in Refs.~\cite{Weickgenannt:2022zxs,Weickgenannt:2022qvh} by the last two terms in each equation, which vanish in Refs.~\cite{Weickgenannt:2022zxs,Weickgenannt:2022qvh} due to the matching condition \eqref{landauJ}.

In order to obtain a closed system of equations for $\omega_0^\mu$ and $\kappa_0^\mu$, we replace the independent spin moments in \eqs\eqref{omegadot2} and \eqref{kappadot2} with their the first-order approximations of the equations of motion derived in Refs.~\cite{Weickgenannt:2022zxs,Weickgenannt:2022qvh} and for convenience again shown in Appendix \ref{spinmomapp} of this work. We also note that spin moments which are parallel to $u^\mu$ in the first index can be expressed through orthogonal ones, see Ref.~\cite{Weickgenannt:2022zxs} and Appendix \ref{spinmomapp} for details.
On the other hand, the new components $u_{[\alpha} \tau_{2\beta]}+\tau_{1[\beta,\alpha]}$ appearing in \eqs\eqref{omegadot2} and \eqref{kappadot2}, which would be zero for Landau matching \eqref{landauS}, cannot be determined from the equations of motion for the corresponding spin moments, since the latter are redundant with the conservation of the spin tensor \eqref{partials}. Instead, they have to be determined through matching conditions. The matching conditions serve to define the spin potential and in general can have a form analogous to the matching conditions for the energy-momentum tensor in Ref.~\cite{Bemfica:2017wps},
\begin{equation}
    u_\lambda \left(S_\Sigma^{\lambda,\mu\nu}-S^{\lambda,\mu\nu}_{\Sigma,\text{eq}}\right)=\mathcal{D}^{\mu\nu}\; , \label{genmatchs}
\end{equation}
where $\mathcal{D}^{\mu\nu}$ is a function of $\beta$, $\alpha$, $u_\mu$, $\Omega_{\mu\nu}$ and their first-order derivatives. In principle, the most general form of $\mathcal{D}^{\mu\nu}$ may be written as a sum of all possible tensor structures at our disposal.
 However, due to the dependence on the spin potential, which is absent in spinless hydrodynamics, in our case taking into account all possible tensor structures would result in long expressions, containing a large number of arbitrary matching coefficients. In order to keep the discussion as simple as possible, we note that causality and stability of the equations of motion for the spin potential require the presence of second-order time derivatives. Hence, we choose matching conditions such that the spin moments appearing in \eq\eqref{genmatchs} are proportional to first-order time derivatives of the spin potential. Inserting \eqs\eqref{spintenssigma} 
 and \eqref{distrnoneqfin} into \eq\eqref{genmatchs}, performing the integrals with the help of the orthogonality relation \eqref{orthrel_poly}, and contracting the result with $ \epsilon_{\alpha\beta\mu\nu} u^\beta$ or $\epsilon_{\alpha\beta\mu\nu}\proj^\alpha_\lambda\proj^\beta_\rho$, respectively, we then find that the following matching conditions are convenient,
\begin{align}
\tau_1^{[\langle\mu\rangle,\nu]}&=\frac{\hbar}{ m}\left(\zeta\, \epsilon^{\mu\nu\alpha\beta}u_\alpha \dot{\kappa}_{0\beta}+\xi\proj^{[\mu}_\lambda \nabla^{\nu]}\omega_0^\lambda\right)\; ,\n\\
\tau_2^{\langle\mu\rangle}-u_\nu \tau_1^{\nu,\mu}&= \frac\hbar m\iota\, \dot{\omega}_0^\lmur\; ,
    \label{minmatch}
\end{align}
where the matching coefficients $\zeta$, $\xi$, and $\iota$ are functions of $\beta$ and $\alpha$. The last term in the first line will serve to simplify the equation of motion for $\kappa_0^\mu$ with an appropriate choice of $\xi$. Since we choose a special form of the matching conditions, the requirements on the transport coefficients derived in the remainder of this paper correspond to sufficient, but not necessary conditions for linear stability and causality. Physically, the matching conditions \eqref{minmatch} relate the change in the spin potential to the dissipative parts of the spin-diffusion tensor
\begin{equation}
\mathfrak{H}^{\nu\mu}\equiv -\frac{1}{2m} \left\langle E_p p^{\langle\nu\rangle}\ms^\mu\right\rangle
\end{equation}
and the spin-energy tensor  
\begin{equation}
\mathfrak{N}^{\nu\mu}\equiv -\frac{1}{2m} u^\nu \left\langle E_p^2 \ms^\mu \right\rangle\; ,
\end{equation}
which are proportional to 
\begin{equation}
\tau_1^{\mu,\nu}\equiv \left\langle E_p p^{\langle\nu\rangle}\ms^\mu\right\rangle-\left\langle E_p p^{\langle\nu\rangle}\ms^\mu\right\rangle_\text{eq}
\end{equation}
and 
\begin{equation}
\tau_2^\mu\equiv \left\langle E_p^2 \ms^\mu \right\rangle-\left\langle E_p^2 \ms^\mu \right\rangle_\text{eq}\; ,
\end{equation}
respectively.

Inserting \eqs\eqref{minmatch} together with the first-order expressions shown in Appendix~\ref{spinmomapp}, \eqs\eqref{taumu}, \eqref{tauparallel24b}, and \eqref{negative}, for all appearing spin moments into \eqs\eqref{omegadot2} and \eqref{kappadot2}, one obtains a closed system of equations of motion for the spin potential. In practice, these equations of motion should be solved in order to determine the spin potential. Then, one obtains an approximate expression for the distribution function by inserting the result for the spin potential into \eqs\eqref{taumu}, which are in turn inserted into \eq\eqref{distrnoneqfin}. From the distribution function one can then calculate all quantities of interest, e.g., the spin tensor or the polarization. 

Due to the length of the equations of motion for the spin potential, we refrain from writing them out here. Instead, we consider a simplified situation, given by small perturbations of a homogeneous, i.e., nonrotating and unpolarized, global equilibrium state,
\begin{align}
    u^\mu &= u^{\mu}_0 +\delta u^\mu\; , &
    \beta&= \beta_0+\delta \beta\; , &
    \alpha&= \alpha_0+ \delta \alpha\; ,\n\\
    \kappa_0^\mu&= \delta \kappa_0^\mu\; ,&
    \omega_0^\mu&= \delta\omega_0^\mu\; .
    \label{deltas}
\end{align}
In the following, all projectors are meant with respect to the unperturbed fluid velocity $u_0^\mu$. Keeping only terms linear in perturbations in \eqs\eqref{omegadot2} and \eqref{kappadot2}, we obtain
\begin{equation}
\dot{\omega}_0^{\langle\alpha\rangle}+\mathfrak{c}_0\ddot{\omega}_0^{\langle\alpha\rangle}=-\mathfrak{c}_1 \nabla^\alpha \nabla\cdot \omega_0-\mathfrak{c}_2 \nabla\cdot\nabla \omega_0^\alpha+\mathfrak{c}_3 \epsilon^{\alpha\lambda\mu\nu}
u_\mu\nabla_\lambda\kappa_{0\nu} \label{omegadot3}
\end{equation}
and
\begin{equation}
\dot{\kappa}_0^{\lmur}+\mathfrak{d}_0\ddot{\kappa}_0^{\langle\mu\rangle}= -\mathfrak{d}_1 \nabla^\mu \nabla\cdot\kappa_0-\mathfrak{d}_2 \nabla\cdot\nabla \kappa_0^\mu+ \mathfrak{d}_3 \epsilon^{\mu\alpha\nu\beta} u_\nu  \nabla_\alpha \omega_{0\beta}\;  ,\label{kappadot4}
\end{equation}
where the calculation and the coefficients in the 14+24-moment approximation are shown in Appendix \ref{coeffapp}. We see that in the linear regime the equations of motion for $\omega_0^\mu$ and $\kappa_0^\mu$ couple to those for the spin-independent quantities only through the transport coefficients, which are functions of $\beta$ and $\alpha$. On the other hand, the equations of motion for $\beta$, $\alpha$, and $u^\mu$ do not depend on the spin potential at all~\cite{Weickgenannt:2022zxs}.~\footnote{In principle, one could introduce a dependence on the spin potential through the matching conditions. However, since this would significantly complicate the calculation without being particularly useful or physically reasonable, we exclude such a matching in this work.} Therefore, it is possible to first solve the equations of motion for conventional BDNK hydrodynamics, and then study spin effects by solving the equations of motion for the spin potential on top. Since stability and causality of the first-order equations of motion for the spin-independent quantities have already been studied in several works~\cite{Bemfica:2017wps,Bemfica:2019knx,Kovtun:2019hdm,Bemfica:2020zjp,Hoult:2020eho}, we will only discuss the equations of motion \eqref{omegadot3} and \eqref{kappadot4} for the spin potential in the following. Note that the decoupling of the equations of motion implies that for the energy-momentum tensor any matching conditions compatible with BDNK hydrodynamics may be chosen. The discussion of the causality and stability of the spin modes in the remainder of this paper is independent of this choice.

We close this section with a remark on the pseudo-gauge choice. The HW pseudo-gauge is particularly convenient for the derivation of first-order spin hydrodynamics, since it directly relates the conservation laws for the energy-momentum tensor and the spin tensor, \eqs\eqref{partialt} and \eqref{partials}, to the microscopic collisional invariants, given by the four-momentum and the dipole-moment tensor, respectively,~\cite{Weickgenannt:2020aaf}
\begin{align}
 \partial_\lambda T^{\mu\lambda} &= \int d\Gamma\, p^\mu \mC[f]=0\; ,\n\\
 \partial_\lambda S^{\lambda,\mu\nu}&= \int d\Gamma\, \Sigma_\ms^{\mu\nu} \mC[f]=0\; .
\end{align}
The presence of these collisional invariants causes zero modes of the collision term, and their equations of motion have to be excluded from the collision matrix when inverting it, as outlined in Appendix \ref{spinmomapp}. Imposing a matching condition on the HW spin tensor, see \eq\eqref{genmatchs}, directly determines the spin moments corresponding to collisional invariants. On the other hand, it would in principle also be possible to choose a matching condition on the total angular-momentum tensor in a different pseudo-gauge, which is always conserved. Due to the arbitrariness of the matching conditions, one could adjust them in a way that \eqs\eqref{minmatch} are recovered. In contrast to second-order spin hydrodynamics, where the choice of dynamical spin moments depends on the pseudo-gauge~\cite{Weickgenannt:2022qvh}, first-order spin hydrodynamics can therefore be derived independently of the pseudo-gauge.

\section{Linear mode expansion}
\label{modesec}

We will now study the stability of small perturbations of the homogeneous global-equilibrium state, see, e.g., Refs.~\cite{Pu:2009fj,Kovtun:2019hdm,Brito:2020nou}. Consider the local rest frame of the unperturbed fluid, $u_{0}^\mu=(1,\mathbf{0})$ and perturbations of the spin potential expanded in linear waves in the fluid rest frame,
\begin{equation}
    \delta \omega_0^\mu = e^{i\Omega t-i\mathbf{k}\cdot \mathbf{x}} (0, \boldsymbol{\omega}_1)\, , \qquad \delta \kappa_0^\mu=e^{i\Omega t-i\mathbf{k}\cdot \mathbf{x}}(0,\boldsymbol{\kappa}_1) \, .
\end{equation}
We can divide any three-vector into components parallel and orthogonal to $\mathbf{k}$,
\begin{equation}
\mathbf{a}=a_{\parallel}\hat{\mathbf{k}}+\mathbf{a}_\bot \; ,
\end{equation}
where we defined $\hat{\mathbf{k}}\equiv\mathbf{k}/k$ with $k\equiv|\mathbf{k}|$. The longitudinal spin modes are obtained by projecting \eqs\eqref{omegadot3} and \eqref{kappadot4} parallel to $\mathbf{k}$. Defining the vector $X_\parallel\equiv (\omega_{1\parallel},\kappa_{1\parallel})$  we find
\begin{equation}
    A_\parallel X_\parallel=0
\end{equation}
with
\begin{equation}
    A_\parallel  \equiv \begin{pmatrix}
    -i\Omega-(\mathfrak{c}_1+\mathfrak{c}_2) k^2+\mathfrak{c}_0 \Omega^2 & 0 \\
    0& -i\Omega-(\mathfrak{d}_1+\mathfrak{d}_2) k^2+\mathfrak{d}_0 \Omega^2
    \end{pmatrix}\; .\label{apar}
\end{equation}
The dispersion relations corresponding to nontrivial solutions $\Omega(k)$ are obtained from
\begin{equation}
    \det A_\parallel=0\; ,
\end{equation}
or equivalently,
\begin{equation}
    (-i\Omega-(\mathfrak{c}_1+\mathfrak{c}_2) k^{2}+\mathfrak{c}_0 \Omega^{2})(-i\Omega-(\mathfrak{d}_1+\mathfrak{d}_2) k^{2}+\mathfrak{d}_0 \Omega^{2})=0 \; .
    \label{rfpm}
\end{equation}
On the other hand, for the transverse spin modes, we project \eqs\eqref{omegadot3} and \eqref{kappadot4} orthogonal to $\mathbf{k}$. With the definition
\begin{equation}
    X_\bot\equiv ( \boldsymbol{\omega}_{1\bot},\boldsymbol{\kappa}_{1\bot})
\end{equation}
we obtain
\begin{equation}
    A_\bot X_\bot =0 \; ,
\end{equation}
where we introduced the $6\times6$ matrix
\begin{equation}
    A_\bot \equiv \begin{pmatrix}
     -i\Omega+\mathfrak{c}_0\Omega^2-\mathfrak{c}_2 k^2 & -i\mathfrak{c}_3 B_\mathbf{k}\\
      -i\mathfrak{d}_3 B_\mathbf{k} & -i\Omega+\mathfrak{d}_0\Omega^2-\mathfrak{d}_2 k^2
    \end{pmatrix} \; , \label{aorth}
\end{equation}
with
\begin{equation}
  B_\mathbf{k}^{ik}\equiv \epsilon^{ijk}k^j \; . 
\end{equation}
In order to obtain the dispersion relations, 
it is more convenient 
to calculate the solution directly than the determinante. We have
\begin{equation}
    -\frac{k^2\mathfrak{d}_3\mathfrak{c}_3}{-i\Omega+\mathfrak{c}_0\Omega^2-\mathfrak{c}_2 k^2} -i\Omega+\mathfrak{d}_0\Omega^2-\mathfrak{d}_2 k^2=0
\end{equation}
and therefore
\begin{equation}
\mathfrak{c}_0\mathfrak{d}_0\Omega^4 -i(\mathfrak{c}_0+\mathfrak{d}_0)\Omega^3-(1+\mathfrak{c}_{(0} \mathfrak{d}_{2)}k^2 )\Omega^2+(i\mathfrak{d}_2k^2+i\mathfrak{c}_2k^2)\Omega+\mathfrak{c}_2 \mathfrak{d}_2 k^4-\mathfrak{c}_3\mathfrak{d}_3 k^2 =0\; . \label{om4rest}
\end{equation}
From the dispersion relations obtained as solutions of \eqs\eqref{rfpm} and \eqref{om4rest}, we will derive conditions for the causality and stability of the theory in the rest frame. We apply the following criteria~\cite{Pu:2009fj,Kovtun:2019hdm,Brito:2020nou}
\begin{align}
 \text{Im}\, \Omega(k) &\geq 0 & \text{(stability)}  \;, \n\\
 \lim_{k\rightarrow\infty} \left|\text{Re}\, \frac{\partial\Omega(k)}{\partial k}\right|&<1 & \text{ (causality)} \; . \label{conditions}
\end{align}
The stability condition guarantees that the spin modes are damped, in contrast a negative imaginary part of $\Omega$ would lead to exponentially growing modes. Furthermore, the causality condition ensures that the group velocity of the spin modes is smaller than the speed of light.
In general, if \eqs\eqref{conditions} are fulfilled in the rest frame, this does not imply causality and stability in an arbitrary frame. However, as we will see in the next sections, in our case we will be able prove stability and causality in any frame  only from the analysis in the rest frame. For the sake of completeness, the analogues of \eqs\eqref{rfpm} and \eq\eqref{om4rest} in a boosted frame and some special solutions for the longitudinal spin modes are shown in Appendix \ref{movingapp}.

\section{Stability and causality conditions for the longitudinal spin modes}
\label{longsec}

\subsection{Stability}

Consider the longitudinal spin modes in the rest frame with the dispersion relations determined by \eq\eqref{rfpm}. We see immediately that if $\mathfrak{c}_0=0$ or $\mathfrak{d}_0=0$, at least one of the dispersion relations has the same form as in Navier-Stokes theory, with $\Omega\sim ik^2$. These types of solutions are never stable in a moving frame. Therefore, we exclude this case and assume $\mathfrak{c}_0\neq 0$, $\mathfrak{d}_0\neq 0$ in the following. The solutions for \eq\eqref{rfpm} then read
\begin{equation}
\Omega_{\parallel1}=\frac i2 \frac{1}{\mathfrak{c}_0}\pm \sqrt{-\frac{1}{4\mathfrak{c}^2_0}+\frac{\mathfrak{c_1}+\mathfrak{c}_2}{\mathfrak{c}_0} k^2}\; , \label{oolong}
\end{equation}
and
\begin{equation}
\Omega_{\parallel2}=\frac i2 \frac{1}{\mathfrak{d}_0}\pm \sqrt{-\frac{1}{4\mathfrak{d}^2_0}+\frac{\mathfrak{d_1}+\mathfrak{d}_2}{\mathfrak{d}_0} k^2}\; .
    \label{solomrfl}
\end{equation}
Note that the dispersion relations depend on the matching coefficients $\zeta$ and $\iota$ through $\mathfrak{c}_0$ and $\mathfrak{d}_0$. Let is first discuss the limit of small wave number for \eq\eqref{oolong}. The two solutions of \eq\eqref{oolong} correspond to a hydrodynamic and a nonhydrodynamic longitudinal spin mode, with the latter having the frequency
\begin{equation}
\Omega_{\parallel1,\text{nh}}\rightarrow \frac{i}{\mathfrak{c}_0} \qquad \qquad (k\rightarrow 0)\; .
\end{equation}
For the nonhydrodynamic modes to be stable in the limit $k\rightarrow 0$, we thus require $\mathfrak{c}_0>0$.  

Considering \eq\eqref{oolong} for arbitrary wave number, we find that for $(\mathfrak{c}_1+\mathfrak{c}_2)/\mathfrak{c}_0>0$ there exists a matching-dependent critical wave number 
\begin{equation}
k_{c} \equiv  \frac{1}{2\sqrt{\mathfrak{c}_0(\mathfrak{c}_1+\mathfrak{c}_2)}} \; ,  
\end{equation}
where for $k>k_c$ the modes propagate.
In order for the longitudinal spin modes to be stable, we need Im $\Omega_\parallel\geq0$, i.e.,
\begin{align}
    \mathfrak{c}_0 > 0 \qquad \qquad &\text{(propagating)}\n\\
    \frac{1}{2\mathfrak{c}_0}-\sqrt{\frac{1}{4\mathfrak{c}^2_0}-\frac{\mathfrak{c_1}+\mathfrak{c}_2}{\mathfrak{c}_0} k^2}>0 \qquad \qquad &\text{(nonpropagating)}
\end{align}
Apparently the stability condition depends on the matching. Inserting $k>k_c$ into the second inequality, we have
\begin{equation}
    \frac{1}{2\mathfrak{c}_0}\left(1-\sqrt{1-4\mathfrak{c}_0(\mathfrak{c}_1+\mathfrak{c}_2)k^2}\right)>\frac{1}{2\mathfrak{c}_0}\left(1-\sqrt{1-4\mathfrak{c}_0(\mathfrak{c}_1+\mathfrak{c}_2)k_c^2}\right)=\frac{1}{2\mathfrak{c}_0}>0\; , \qquad \qquad ((\mathfrak{c}_1+\mathfrak{c}_2)/\mathfrak{c}_0>0)\; .
\end{equation}
Therefore, if $\mathfrak{c}_1+\mathfrak{c}_2\geq 0$, all longitudinal spin modes are stable in the rest frame if and only if $\mathfrak{c}_0>0$. If $\mathfrak{c}_1+\mathfrak{c}_2<0$ and $\mathfrak{c}_0<0$, there are unstable propagating modes. If $\mathfrak{c}_1+\mathfrak{c}_2<0$ and $\mathfrak{c}_0>0$, there are unstable nonpropagating modes. 

An analogous discussion can be carried out for \eq\eqref{solomrfl}. Hence, the stability conditions for the longitudinal spin modes in the rest frame can be summarized as
\begin{align}
    \mathfrak{c}_0& > 0 \; , & \mathfrak{d}_0& > 0 \; , \n\\
   \mathfrak{c}_1+\mathfrak{c}_2& \geq 0 \; , & \mathfrak{d}_1+\mathfrak{d}_2& \geq 0\; .
   \label{cond1}
\end{align}

\subsection{Causality}

Next, we discuss the requirements from the causality of the longitudinal spin modes in the rest frame. The group velocity $v_g\equiv \text{Re }\partial\Omega/\partial k$ of the propagating modes corresponding to the solutions \eqref{oolong} is determined as
\begin{equation}
   v_{g\parallel1}= \frac{(\mathfrak{c}_1+\mathfrak{c}_2)k}{\mathfrak{c}_0\sqrt{-\frac{1}{4\mathfrak{c}^2_0}+\frac{\mathfrak{c_1}+\mathfrak{c}_2}{\mathfrak{c}_0} k^2}} \; .
\end{equation}
For $k\rightarrow\infty$ the group velocity becomes
\begin{equation}
v_{g\parallel1}=\sqrt{\left|\frac{\mathfrak{c}_1+\mathfrak{c}_2}{\mathfrak{c}_0}\right|} \; ,\qquad \qquad (k\rightarrow\infty)\; , \label{vgompar}
\end{equation}
hence we need 
\begin{equation}
|\mathfrak{c}_0|>|\mathfrak{c}_1+\mathfrak{c}_2|\; \label{cond2}
\end{equation}
for causality. An analogous discussion of \eq\eqref{solomrfl} yields the condition
\begin{equation}
|\mathfrak{d}_0|>|\mathfrak{d}_1+\mathfrak{d}_2|\; . \label{cond2b}
\end{equation}
Equations \eqref{cond2} and \eqref{cond2b} give a minimal required absolute value of the matching coefficients $\zeta$ and $\iota$. 

As mentioned before, in general, causality and stability in the rest frame do not imply causality and stability in a moving frame. However, here we have that $\Omega\sim k$ for $k\rightarrow\infty$. In this case, the causality in the moving frame can directly be proven from \eq\eqref{vgompar}. Since the group velocity is constant, the boosted group velocity can be obtained from \eq\eqref{vgompar} by simply applying relativistic addition of velocities, and hence it cannot become larger than 1~\cite{Kovtun:2019hdm}. Furthermore, if the theory is causal in any frame and stable in the rest frame, it is also stable in any frame~\cite{Bemfica:2020zjp}. Since the calculations for the longitudinal spin modes in a boosted frame are still compact, the causality conditions in a moving frame are for illustration purposes shown to be identical to those in the rest frame in Appendix \ref{movingapp}.

\section{Stability and causality conditions for the transverse spin modes}
\label{transsec}

\subsection{Stability}

Let us now discuss the transverse spin modes with the dispersion relations given by the solutions of \eq\eqref{om4rest}. Again we first consider the limit of small wave number.
  The nonhydrodynamic modes for $k\rightarrow 0$ have the frequency
\begin{equation}
\Omega_{\bot1,\text{nh}}=\frac{i}{\mathfrak{d}_0}\; , \qquad \Omega_{\bot2,\text{nh}}=\frac{i}{\mathfrak{c}_0}\; , \qquad \qquad (k\rightarrow 0)\; .
\end{equation}
These modes are nonpropagating and stable in this limit due to the already known conditions \eqref{cond1}. Furthermore, we have for the hydrodynamic modes ($\Omega\sim k$ for $k\rightarrow0$)
\begin{equation}
    \Omega_{\bot,\text{h}} = \pm i \sqrt{\mathfrak{c}_3\mathfrak{d}_3} k\; , \qquad \qquad (k\rightarrow 0)\; .
\end{equation}
We therefore obtain another stability condition,
\begin{equation}
\mathfrak{c}_3\mathfrak{d}_3<0 \; .
\end{equation}

The solution of \eq\eqref{om4rest} for arbitrary wave number can be straightforwardly obtained, however, we refrain from writing it out due to the length of the expression. Instead, in order to investigate stability, we rewrite \eq\eqref{om4rest} with $\Omega_\bot\equiv-i\bar{\Omega}_\bot$ as
\begin{equation}
\mathfrak{c}_0\mathfrak{d}_0\bar{\Omega}_\bot^4 +(\mathfrak{c}_0+\mathfrak{d}_0)\bar{\Omega}_\bot^3+(1+\mathfrak{c}_{(0} \mathfrak{d}_{2)}k^2 )\bar{\Omega}_\bot^2+(\mathfrak{d}_2+\mathfrak{c}_2)k^2\bar{\Omega}_\bot+\mathfrak{c}_2 \mathfrak{d}_2 k^4-\mathfrak{c}_3\mathfrak{d}_3 k^2 =0\; . \label{om4restspec2}
\end{equation}
Since this is a fourth-order polynomial in $\bar{\Omega}_\bot$ with real coefficients, we may apply the Routh-Hurwitz criterion for the stability condition Re$\, \bar{\Omega}_\bot<0$, see, e.g., Ref.~\cite{Kovtun:2019hdm}. Assuming the conditions \eqref{cond1} to be fulfilled, all coefficients are larger than zero for all $k$ if in addition
\begin{equation}
\mathfrak{c}_2\mathfrak{d}_2 >0 \; , \qquad \qquad \mathfrak{c}_3\mathfrak{d}_3 <0 \; .
\label{cond3}
\end{equation}
The Routh-Hurwitz criterion yields the additional requirement
\begin{equation}
    \frac{\mathfrak{c}_2 \mathfrak{d}_2 k^4-\mathfrak{c}_3\mathfrak{d}_3 k^2}{\mathfrak{c}_0\mathfrak{d}_0}< \frac{(\mathfrak{d}_2+\mathfrak{c}_2)k^2}{\mathfrak{c}_0+\mathfrak{d}_0}\left(\frac{1+\mathfrak{c}_{(0} \mathfrak{d}_{2)}k^2}{\mathfrak{c}_0\mathfrak{d}_0}-\frac{(\mathfrak{d}_2+\mathfrak{c}_2)k^2}{\mathfrak{c}_0+\mathfrak{d}_0} \right)\; .
\end{equation}
The inequality has to hold for any $k$. Considering the terms $\sim k^4$, the relation can be simplified to 
\begin{equation}
    (\mathfrak{c}_0\mathfrak{d}_2-\mathfrak{c}_2\mathfrak{d}_0)^2 k^4 >0\; ,
\end{equation}
which is always fulfilled, since the coefficients are real, and does not result in an additional condition.
On the other hand, from the terms $\sim k^2$
, we obtain the last stability condition
\begin{equation}
    -\mathfrak{c}_3 \mathfrak{d}_3 < \frac{\mathfrak{c}_2+\mathfrak{d}_2}{\mathfrak{c}_0+\mathfrak{d}_0}\; . \label{cond4}
\end{equation}
Given that \eqs\eqref{cond3} and \eqref{cond4} are fulfilled, the transverse spin modes are stable in the rest frame. 

\subsection{Causality}

The solutions of \eq\eqref{om4rest} for $k\rightarrow\infty$ read
\begin{align}
    \Omega_{\bot1}^2&= \frac{\mathfrak{d}_2}{\mathfrak{d}_0}k^2\; , \qquad \qquad \Omega_{\bot2}^2= \frac{\mathfrak{c}_2}{\mathfrak{c}_0}k^2\; , \qquad \qquad (k\rightarrow \infty) \; .
\end{align}
Hence, the group velocities are obtained as
\begin{equation}
    v_{g\bot1}= \text{Re } \sqrt{\frac{\mathfrak{d}_2}{\mathfrak{d}_0}}\; , \qquad \qquad v_{g\bot2}=\text{Re } \sqrt{\frac{\mathfrak{c}_2}{\mathfrak{c}_0}}\;, \qquad \qquad (k\rightarrow \infty)\; .
\end{equation}
Since the terms under the square roots are positive, this yields the causality conditions
\begin{equation}
    |\mathfrak{d}_2|<|\mathfrak{d}_0|\; , \qquad \qquad |\mathfrak{c}_2|<|\mathfrak{c}_0|\; .
    \label{caustrans}
\end{equation}
However, these requirements are already implied by the previously found conditions \eqref{cond1}, \eqref{cond2}, \eqref{cond2b}, and \eqref{cond3}. As $\Omega\sim k$ for $k\rightarrow\infty$, the group velocity is again constant and causality in the rest frame implies causality in any frame. Together with the conditions for stability in the rest frame, stability in any frame in guaranteed.

\section{Conclusions}
\label{concsec}

In this paper, we derived linearly stable and causal equations of motion for first-order spin hydrodynamics from kinetic theory. The crucial step is to choose matching conditions which relate the spin energy and spin diffusion to time derivatives of the spin potential. This introduces second-order time derivatives in the equations of motion for the spin potential, rendering them  causal and stable for a certain class of transport coefficients, despite the presence second-order spatial derivatives, analogously to spinless BDNK hydrodynamics. Linearizing the equations of motion for the spin potential around homogeneous global equilibrium, we obtained the compact equations of motion \eqref{omegadot3} and \eqref{kappadot4}. We then derived conditions for causality and stability of these equations in any Lorentz frame by analyzing the dispersion relations for the linear modes corresponding to small perturbations of the spin potential. 
These conditions on the transport coefficients are summarized as follows,
\begin{align}
 \mathfrak{c}_0>  \mathfrak{c}_1+\mathfrak{c}_2& \geq 0 \; , & \mathfrak{d}_0>\mathfrak{d}_1+\mathfrak{d}_2& \geq 0\; ,\n\\ 
 \mathfrak{c}_2\mathfrak{d}_2 &>0 \; , &\mathfrak{c}_3\mathfrak{d}_3 &<0 \; ,\n\\
 -\mathfrak{c}_3 \mathfrak{d}_3 &< \frac{\mathfrak{c}_2+\mathfrak{d}_2}{\mathfrak{c}_0+\mathfrak{d}_0}\; .
 \label{finalcond}
\end{align}
Note that with our choice of matching conditions only $\mathfrak{c}_0$ and $\mathfrak{d}_0$ depend on the matching coefficients $\iota$ and $\zeta$. The other transport coefficients are fixed functions of temperature and chemical potential, some of them also depending on the microscopic interaction. Therefore, one has to check whether the transport coefficients fulfill all requirements, potentially depending on the model used for the collision term, before applying the first-order spin hydrodynamics derived in this work. Even if the stability and causality conditions \eqref{finalcond} are not fulfilled in a particular situation, this does not imply that first-order spin hydrodynamics cannot be causal and stable in that case. Instead, one may try to choose more general matching conditions and adjust the additional matching coefficients accordingly, at the expense of facing more complicated equations of motion. On the other hand, if all conditions \eqref{finalcond} are fulfilled for some choice of $\iota$ and $\zeta$, the matching conditions employed in this work are sufficient for linear stability and causality of first-order spin hydrodynamics. 

The theory of causal and stable first-order spin hydrodynamics derived in this paper has potential applications both for relativistic heavy-ion collisions and astrophysics. In particular, it may be used to dynamically calculate the local Lambda polarization for heavy-ion collisions. If the causality and stability conditions are fulfilled, one can numerically solve the equations of motion, obtaining a result for the spin potential, and hence for the spin tensor and the polarization, valid up to first order in gradients. It would be interesting to compare such a result both to local-equilibrium calculations~\cite{Liu:2021uhn,Fu:2021pok,Becattini:2021suc,Becattini:2021iol} and results from second-order spin hydrodynamics derived in Refs.~\cite{Weickgenannt:2022zxs,Weickgenannt:2022qvh}. This may shed light on the importance of polarization dynamics beyond local equilibrium in heavy-ion collisions.
Furthermore, one may extend the work done in this paper by analyzing linear stability conditions around an inhomogeneous (polarized and/or rotating) equilibrium state, and general causality conditions not restricted to the linear regime~\cite{Bemfica:2019knx}. As another possible extension, one may also consider a nonlocal collision term, which is responsible for the conversion between orbital angular momentum and spin, and will therefore lead to a coupling between the equations of motion for the spin potential and the other thermodynamic potentials.

\section*{Acknowledgements}

I thank Y.\ Peng, D.\ H.\ Rischke, J.\ Sammet, M.\ Shokri,  E.\ Speranza, and D.\ Wagner for enlightening discussions. Furthermore, I thank J.-P.\ Blaizot for valuable comments on the manuscript. N.W.\ acknowledges support by the German National Academy of Sciences Leopoldina through the Leopoldina fellowship program with funding code LPDS 2022-11.

\begin{appendix}

\section{First-order spin moments from equations of motion}
\label{spinmomapp}

In this appendix, we shortly outline how to obtain the expressions for the spin moments valid up to first order in a gradient expansion, which we will insert into \eqs\eqref{omegadot3} and \eqref{kappadot4}. The steps are analogous to those performed in Ref.~\cite{Weickgenannt:2022zxs} and make use of the same assumptions, and we refer to that work for a more detailed discussion. Keeping only first-order terms in the equations of motion for the spin moments obtained in Ref.~\cite{Weickgenannt:2022zxs}, we find
\begin{align}
-\mC^{\langle\mu\rangle}_{r-1}=& 
\frac{\hbar}{2m}\xi_r^{(0)} \theta\, \omega_0^\mu 
-\frac{\hbar}{4m} I_{(r+1)1}\proj^\mu_\lambda\nabla_\nu \tilde{\Omega}^{\lambda\nu}-\frac{\hbar}{4m} \tilde{\Omega}^{\langle\mu\rangle\nu}  
 I_{(r+1)1} I_\nu -\frac{\hbar}{4m} I_{(r+1)0} \epsilon^{\mu\nu\alpha\beta} u_\nu \dot{\Omega}_{\alpha\beta} \;,\n\\
 -\mC^{\langle\mu\rangle,\langle\nu\rangle}_{r-1}
=&\frac{\hbar}{4m} \proj^\mu_\rho\proj^\nu_\lambda \tilde{\Omega}^{\rho\lambda}\xi_r^{(1)}\theta +\frac{\hbar}{4m}\proj^\mu_\rho\proj^\nu_\lambda \dot{\tilde{\Omega}}^{\rho\lambda}I_{(r+2)1}-\frac{\hbar}{2m} \omega_0^\mu I_{(r+2)1} I^\nu-\frac{\hbar}{2m}\beta I_{(r+3)2}
\tilde{\Omega}^{\langle\mu\rangle}_{\ \lambda} \sigma^{\nu\lambda}\n\\
&-\frac{\hbar}{4m} I_{(r+2)1}\proj^\mu_\rho 
\left(\nabla^\nu\tilde{\Omega}^{\rho\lambda}\right)u_\lambda\; ,\n\\
-\mC_{r-1}^{\lmur,\langle\nu\lambda\rangle}
=& \frac{\hbar}{2m} \xi_r^{(2)} \tilde{\Omega}^{\lmur\langle\nu}I^{\lambda\rangle}
+\frac{\hbar}{2m} I_{(r+3)2} \proj^\mu_\rho\proj^{\nu\lambda}_{\alpha\beta} 
\nabla^\alpha \tilde{\Omega}^{\rho\beta}-\frac{\hbar}{2m}\tilde{\Omega}^{\mu\rho}\beta u_\rho 
\sigma^{\nu\lambda} I_{(r+4)2}
\label{eomtaumu}
\end{align}
with 
\begin{align}
\xi_r^{(0)}&\equiv - I_{(r+1)0}-r\, I_{(r+1)1}
-\frac{1}{D_{20}}\left[G_{2(r+1)}(\epsilon_0+P_0) -G_{3(r+1)}n_0\right] \; ,\n\\
 \xi_r^{(1)}&\equiv \frac{G_{3(r+2)}}{D_{20}}n_0-\frac{G_{2(r+2)}}{D_{20}}(\epsilon_0+P_0)-\frac53 \beta I_{(r+3)2}\; ,\n\\
 \xi_r^{(2)}&\equiv I_{(r+3)2}-\frac{n_0}{\epsilon_0+P_0} I_{(r+4)2}\;,
\end{align}
where $n_0$ is the local-equilibrium particle density, $\epsilon_0$ is the local-equilibrium energy density, and $P_0$ is the thermodynamic pressure.
Furthermore, we defined $I^\mu\equiv \nabla^\mu \alpha_0$, the thermodynamic functions
\begin{equation}
I_{nq}(\alpha,\beta)\equiv {\frac{1}{(2q+1)!!}} \int d\Gamma\,  E_p^{n-2q}(-\proj^{\alpha\beta} 
p_\alpha p_\beta)^q f_{\text{eq}}(x,p,\ms)\label{thermdynint}
\end{equation}
and
\begin{equation}
G_{nm}\equiv I_{n0} I_{m0}-I_{(n-1)0}I_{(m+1)0}\;, \qquad \qquad
D_{nq}\equiv I_{(n+1)q}I_{(n-1)q}-I_{nq}^2\;,
\end{equation}
as well as the collision integrals
\begin{equation} \label{coll_int}
\mC_r^{\mu,\langle \mu_1\cdots \mu_n\rangle}\equiv \int d\Gamma\, E_p^r\, p^{\langle\mu_1} 
\cdots p^{\mu_n\rangle} \ms^\mu \mC[f]\;.
\end{equation}
The latter can be expressed through spin moments as~\cite{Weickgenannt:2022zxs}
\begin{align}
\mC_{r-1}^\mu&=-\sum_{n \in \mathbb{S}_0} B_{rn}^{(0)} 
\taum_n^{\mu}\;,\n\\
\mC_{r-1}^{\mu,\langle\nu\rangle}&=-\sum_{n \in \mathbb{S}_1} B_{rn}^{(1)} 
\taum_n^{\mu,\langle\nu\rangle} \;, \n\\
\mC_{r-1}^{\mu,\langle\nu \lambda\rangle}&=  -\sum_{n \in \mathbb{S}_2} B_{rn}^{(2)} 
\taum_n^{\mu,\langle\nu \lambda\rangle}
 \;, 
\label{ccbaridontknow}
\end{align}
where
\begin{align}
B_{rn}^{(l)}& \equiv -16 \frac{1}{2l+1} \proj^{\nu_1\cdots\nu_l}_{\mu_1\cdots\mu_l} 
 \int dP\, dP^\prime\, dP_1\, dP_2\, \mathcal{W}_0 f_{\text{eq}} ^{(0)}(x,p)f_{\text{eq}}^{(0)}(x,p^\prime) E_p^{r-1} p^{\langle\mu_1}\cdots p^{\mu_l\rangle}  
\mathcal{H}_{pn}^{(m)}p_{\langle\nu_1}\cdots p_{\nu_m\rangle}\;,\label{bmanymunu}
\end{align}
with $\mathcal{W}_0$ being the transition rate, see Refs.~\cite{Weickgenannt:2021cuo,Weickgenannt:2022zxs} for details. The coefficient function $ \mathcal{H}_{pn}^{(l)}$ is given by
\begin{equation}
\mathcal{H}_{pn}^{(l)}=\frac{w^{(l)}}{l!}\sum_{m=n}^{N_l} a_{mn}^{(l)} \mathcal{P}_{pm}^{(l)}\;, \label{Hcoeffdef}
\end{equation}
with  
\begin{equation}
\mathcal{P}_{pn}^{(l)}\equiv \sum_{r=0}^n a_{nr}^{(l)} E_p^r \label{polynomialE}
\end{equation}
being orthogonal polynomials in energy. The coefficients $a_{nr}^{(l)}$ are 
determined such that
\begin{equation} \label{orthrel_poly}
2\int dP\, \frac{w^{(l)}}{(2l+1)!!}\left(\proj^{\alpha\beta}p_\alpha p_\beta\right)^l f^{(0)}(x,p) 
\mathcal{P}_{pm}^{(l)} \mathcal{P}_{pn}^{(l)}=\delta_{mn}\;,
\end{equation}
and the normalization reads $w^{(l)} = (-1)^l/I_{2l,l}$.
 Note that we neglect contributions from the nonlocal collision term in this work.
For spin moments with non-negative $n$ and $l\neq1$, we invert the matrix $B$ to obtain 
\begin{align}
\taum^{\mu,\mu_1\cdots\mu_l}_n&=-\sum_{r \in \mathbb{S}_l} 
\mathfrak{T}_{nr}^{(l)}\, {\mathfrak{C}}_{r-1}^{\mu,\langle\mu_1\cdots\mu_l\rangle}\;,
\label{invertB}
\end{align}
where we defined the matrix
\begin{equation} \label{TBinv}
\mathfrak{T}^{(l)}\equiv \left( B^{(l)}\right)^{-1}\;.
\end{equation}
For $l=1$, we symmetrize \eq\eqref{invertB} in the two Lorentz indices, since the antisymmetric part of $\tau^{\langle\mu\rangle,\mu_1}_1$, which is the only spin moment of tensor rank two we need in \eqs\eqref{omegadot3} and \eqref{kappadot4}, corresponds to a collisional invariant related to the conservation of the spin tensor and  is determined by the matching conditions. We thus obtain
\begin{align}
\tau^{\langle\mu\rangle}_{n}=& \tc^{(0)}_{\theta\omega,n}
 \theta\, \omega_0^\mu + \tc^{(0)}_{\nabla\Omega,n}
\proj^\mu_\lambda\nabla_\nu \tilde{\Omega}^{\lambda\nu}+ \tc^{(0)}_{I\Omega,n}\tilde{\Omega}^{\langle\mu\rangle\nu}  
 I_\nu+ \tc^{(0)}_{\dot{\Omega},n}\epsilon^{\mu\nu\alpha\beta} u_\nu \dot{\Omega}_{\alpha\beta} \;,\n\\
\tau^{(\langle\mu\rangle,\langle\nu\rangle)}_{n}
=& -\tc^{(1)}_{\omega I,n} \omega_0^{(\mu}  I^{\nu)}+\tc^{(1)}_{\Omega\sigma,n}
\tilde{\Omega}^{(\langle\mu\rangle}_{\ \lambda} \sigma^{\nu)\lambda}-\tc^{(1)}_{\nabla\Omega,n}\proj^{(\mu}_\rho 
\left(\nabla^{\nu)}\tilde{\Omega}^{\rho\lambda}\right)u_\lambda\; ,\n\\
\tau_{n}^{\lmur,\langle\nu\lambda\rangle}
=& \tc^{(2)}_{\Omega I,n}\tilde{\Omega}^{\lmur\langle\nu}I^{\lambda\rangle}
+ \tc^{(2)}_{\nabla\Omega,n} \proj^\mu_\rho\proj^{\nu\lambda}_{\alpha\beta} \nabla^\alpha \tilde{\Omega}^{\rho\beta}-\tc^{(2)}_{\omega\sigma,n}\tilde{\Omega}^{\mu\rho} u_\rho 
\sigma^{\nu\lambda} 
\label{taumu}
\end{align}
with 
\begin{align}
\tc^{(0)}_{\theta\omega,n} &\equiv \frac{\hbar}{2m}\sum_{r \in \mathbb{S}_l} 
\mathfrak{T}_{nr}^{(l)}\xi_r^{(0)}\; ,\n\\
\tc^{(0)}_{\nabla\Omega,n} &\equiv -\frac{\hbar}{4m}\sum_{r \in \mathbb{S}_l} 
\mathfrak{T}_{nr}^{(l)} I_{(r+1)1} \equiv
\tc^{(0)}_{I\Omega,n} \; , \n\\
\tc^{(0)}_{\dot{\Omega},n} &\equiv -\frac{\hbar}{4m}\sum_{r \in \mathbb{S}_l} 
\mathfrak{T}_{nr}^{(l)} I_{(r+1)0} \; ,\n\\
\tc^{(1)}_{\omega I,n} &\equiv \frac{\hbar}{2m}\sum_{r \in \mathbb{S}_l} 
\mathfrak{T}_{nr}^{(l)}I_{(r+2)1} \; ,\n\\
\tc^{(1)}_{\Omega\sigma,n} &\equiv -\frac{\hbar}{2m}\beta\sum_{r \in \mathbb{S}_l} 
\mathfrak{T}_{nr}^{(l)} I_{(r+3)2}  \; , \n\\
\tc^{(1)}_{\nabla\Omega,n}&\equiv \frac{\hbar}{4m} \sum_{r \in \mathbb{S}_l} 
\mathfrak{T}_{nr}^{(l)}I_{(r+2)1}\; ,\n\\
\tc^{(2)}_{\Omega I,n}&\equiv \frac{\hbar}{2m} \sum_{r \in \mathbb{S}_l} 
\mathfrak{T}_{nr}^{(l)}\xi_r^{(2)}  \; ,\n\\
\tc^{(2)}_{\nabla\Omega,n}&\equiv \frac{\hbar}{2m}\sum_{r \in \mathbb{S}_l} 
\mathfrak{T}_{nr}^{(l)} I_{(r+3)2}\; , \n\\
\tc^{(2)}_{\omega\sigma,n} &\equiv \frac{\hbar}{2m}\beta\sum_{r \in \mathbb{S}_l} 
\mathfrak{T}_{nr}^{(l)}  I_{(r+4)2}\; .
\label{transcoeff} 
\end{align}
The minimal set of summation indices is given by the 14+24-moment truncation, $\mathbb{S}_0=\{0,2\}$, $\mathbb{S}_1=\{1\}$, $\mathbb{S}_2=\{0\}$, for which the transport coefficients \eqref{transcoeff} for the respective spin moments reduce to those given in Ref.~\cite{Weickgenannt:2022zxs}. This set may be extended in order to improve the approximation. Note that in any case $r=2$ has to be excluded from $\mathbb{S}_0$ and $r=1$ from $\mathbb{S}_1$ for the antisymmetric spin moments when inverting the collision term, since $\tau_2^{\lmur}$ and $\tau_1^{[\lmur,\nu]}$ are parts of the collisional invariants and fixed by the matching conditions. The components of the spin moments parallel to $u_\mu$ are expressed through the orthogonal ones using~\cite{Weickgenannt:2022zxs}
\begin{align}
u_\mu \taum_r^\mu =& - \taum_{r-1,\mu}^{\mu}\; ,\n\\
u_\mu \taum_r^{\mu,\nu}=& -\taum_{r-1\,\mu}^{\mu,\nu}
-\frac13 \left(m^2 \taum_{r-1}^{\langle\nu\rangle}-\taum_{r+1}^{\langle\nu\rangle}\right)\; ,\n\\
u_\mu \taum_r^{\mu,\nu\lambda}=& -\taum_{r-1\ \mu}^{\mu,\nu\lambda}
+\frac{2}{15} \left(m^2 \taum^\mu_{r-1,\mu}-\taum^\mu_{r+1,\mu} \right)\proj^{\nu\lambda}
-\frac15 \left( m^2 \taum_{r-1}^{(\langle\nu\rangle,\lambda)}
-\taum_{r+1}^{(\langle\nu\rangle,\lambda)}\right)\;. 
\label{tauparallel24b}
\end{align}
Finally, spin moments with negative $r$ are obtained from the relation~\cite{Weickgenannt:2022zxs}
\begin{equation}
\taum_{-r}^{\mu,\mu_1\cdots\mu_l}=  \sum_{n\in \mathbb{S}_l} \taum^{\mu,\mu_1\cdots\mu_l}_n 
\mathfrak{F}_{-rn}^{(l)} \;, \label{negative}
\end{equation}
with
\begin{equation}
\mathfrak{F}_{-rn}^{(l)}\equiv \frac{2l!}{(2l+1)!!} \int dP\, E_p^{-r}\, 
\mathcal{H}_{pn}^{(l)} \left(\proj^{\alpha\beta}p_\alpha p_\beta\right)^{l} f^{(0)}_\text{eq}(x,p)\;,
\end{equation}
see also Ref.~\cite{Denicol:2012cn}.

\section{Transport coefficients for linearized equations of motion}
\label{coeffapp}

In order to derive \eqs\eqref{omegadot3} and \eqref{kappadot4} from \eqs\eqref{omegadot2} and \eqref{kappadot2}, we insert the following spin moments, obtained by linearizing \eqs\eqref{taumu}, 
\begin{align}
\tau^{\langle\mu\rangle}_{0}=& 2\tc^{(0)}_{\nabla\Omega,0} \epsilon^{\mu\nu\alpha\beta}u_\alpha \nabla_\nu \kappa_{0\beta}
+ 2\tc^{(0)}_{\dot{\Omega},0}\dot{\omega}_0^\mu \;,\n\\
\tau^{(\langle\mu\rangle,\nu)}_{1}
=& -2\tc^{(1)}_{\nabla\Omega,1} \nabla^{(\nu}\omega_0^{\mu)}\; ,\n\\
\tau_{0}^{\lmur,\nu\lambda}
=& 2 \tc^{(2)}_{\nabla\Omega,0}u_\tau\epsilon^{\mu\tau\sigma\langle\nu}  \nabla^{\lambda\rangle} \kappa_{0\sigma}\; .
\end{align}
We furthermore find from \eqs\eqref{tauparallel24b}
\begin{align}
    u_\lambda \left(m^2 \tau_{0}^\lambda-\taum_2^\lambda\right)&= 2\left(m^2 \mathfrak{F}^{(1)}_{-11}-1 \right)\tc^{(1)}_{\nabla\Omega,1} \nabla \cdot \omega_0\; ,\n\\
    u_\mu \taum_0^{\mu,\nu\lambda}=& -\frac{4}{15} \left(m^2 \mathfrak{F}^{(1)}_{-11}-1 \right)\proj^{\nu\lambda} \tc^{(1)}_{\nabla\Omega,1} \nabla \cdot \omega_0
+\frac25 \left(m^2 \mathfrak{F}^{(1)}_{-11}-1 \right)\tc^{(1)}_{\nabla\Omega,1} \nabla^{(\nu}\omega_0^{\lambda)}\; ,
\end{align}
where we used the 14+24-moment approximation. 
With the matching conditions \eqref{minmatch} we obtain
\begin{equation}
\tau_2^{\langle\mu\rangle}+\taum_{0\,\ \ \nu}^{\nu,\mu}
+\frac13 \left(m^2 \taum_{0}^{\langle\mu\rangle}-\taum_{2}^{\langle\mu\rangle}\right)= \frac\hbar m\iota\, \dot{\omega}_0^\mu\; ,
\end{equation}
which implies
\begin{align}
    \tau_2^\lmur&=\left(\frac32 \frac\hbar m\iota-m^2\tc^{(0)}_{\dot{\Omega},0}\right) \dot{\omega}_0^\mu-\left(m^2\tc^{(0)}_{\nabla\Omega,0}-\frac52 \tc^{(2)}_{\nabla\Omega,0} 
\right)\epsilon^{\mu\nu\alpha\beta}u_\alpha \nabla_\nu \kappa_{0\beta}\; .
\end{align}
Using these results in \eqs\eqref{omegadot2} and \eqref{kappadot2}, we obtain at linear order in perturbations
\begin{align}
\frac{\hbar}{m^2}\dot{\omega}_0^{\alpha}= 
& -\frac{2}{I_{30}- I_{31}} \bigg\{ \frac{2}{3m}\left(m^2 \mathfrak{F}^{(1)}_{-11}-1 \right)\tc^{(1)}_{\nabla\Omega,1} \nabla^\alpha \nabla \cdot \omega_0
-\frac12\frac{\hbar}{m^2}\epsilon^{\alpha\lambda\mu\nu}I_{31} 
u_\mu\nabla_\lambda\kappa_{0\nu}+\frac{1}{m}  
\tc^{(1)}_{\nabla\Omega,1} \nabla_\lambda \nabla^{(\alpha}\omega_0^{\lambda)} \n\\
&+\frac1m   \left[\frac{2}{15} \left(m^2 \mathfrak{F}^{(1)}_{-11}-1 \right)\nabla^{\alpha} \tc^{(1)}_{\nabla\Omega,1} \nabla \cdot \omega_0
+\frac25 \left(m^2 \mathfrak{F}^{(1)}_{-11}-1 \right)\tc^{(1)}_{\nabla\Omega,1} \nabla\cdot\nabla\omega_0^{\alpha}\right]-\frac{\hbar}{m^2}\iota\, \ddot{\omega}_0^\alpha\bigg\}
\end{align}
and
\begin{align}
\frac{\hbar}{m^2}\dot{\kappa}^{\mu}_0=&  -\frac{1}{I_{31}} \bigg\{ -I_{31} \frac{\hbar}{2m^2}\epsilon^{\alpha\mu\nu\beta} u_\nu  \nabla_\alpha \omega_{0\beta}-{\frac{1}{3m}}\left(3m^2\tc^{(0)}_{\dot{\Omega},0}-\frac32 \frac\hbar m\iota\right) \epsilon^{\alpha\mu\nu\beta}u_\nu\nabla_\alpha \dot{\omega}_{0\beta} 
\n\\
&-{\frac{1}{3m}}\left(3m^2 \tc^{(0)}_{\nabla\Omega,0} -\frac52 \tc^{(2)}_{\nabla\Omega,0} \right)\nabla_\alpha \nabla^{[\mu}\kappa_0^{\alpha]}-\frac{1}{m} \tc^{(2)}_{\nabla\Omega,0}\left(-2\nabla\cdot\nabla\kappa_0^\mu+\frac13\nabla_\alpha \nabla^{[\mu} \kappa_0^{\alpha]}\right)+\frac{\hbar}{ m^2} \zeta \ddot{\kappa}_0^\mu\n\\
&
-\frac{\hbar}{m^2} \xi\epsilon^{\mu\nu\alpha\beta}u_\nu \nabla_{\alpha}\dot{\omega}_{0\beta}\bigg\} \; .
\end{align}
Choosing
\begin{equation}
    \xi= -\frac{m}{3\hbar}\left(3m^2\tc^{(0)}_{\dot{\Omega},0}-\frac32 \frac\hbar m\iota\right)\; ,
\end{equation}
we arrive at the equations of motion \eqref{omegadot3} and \eqref{kappadot4} with the coefficients given by
\begin{align}
    \mathfrak{c}_0&\equiv -\frac{2}{I_{30}-I_{31}}\iota\; , & \mathfrak{c}_1& \equiv \frac{m}{\hbar}\frac{2}{5(I_{30}-I_{31})} \tc^{(1)}_{\nabla\Omega,1}(4m^2\mathfrak{F}_{-11}^{(1)}+1)\;,\n\\
    \mathfrak{c}_2&\equiv \frac{m}{\hbar}\frac{2}{5(I_{30}-I_{31})} \tc^{(1)}_{\nabla\Omega,1}(2m^2\mathfrak{F}_{-11}^{(1)}+3)\; , & \mathfrak{c}_3&\equiv \frac{I_{31}}{I_{30}-I_{31}}
\end{align}
and
\begin{align}
\mathfrak{d}_0&\equiv \frac{1}{ I_{31}}\zeta\; , &     \mathfrak{d}_1&\equiv\frac{m}{\hbar I_{31}}\left(-m^2 \tc^{(0)}_{\nabla\Omega,0}+\frac12\tc^{(2)}_{\nabla\Omega,0}\right)\;, \n\\
\mathfrak{d}_2&\equiv \frac{m}{\hbar}\frac{1}{I_{31}}\left(\frac32\tc^{(2)}_{\nabla\Omega,0}+m^2\tc^{(0)}_{\nabla\Omega,0}\right)\;,&
    \mathfrak{d}_3&\equiv -\frac12\;  .
\end{align}

\section{Dispersion relations in a moving frame}
\label{movingapp}

We consider a general background velocity $u_0^{\mu}\equiv(\gamma,\gamma V, 0, 0)$ with $\gamma\equiv1/\sqrt{1-V^2}$, which can be obtained from a Lorentz boost with velocity $V$ of $u_{0\star}^\mu\equiv(1,\mathbf{0})$, assuming the boost to be in $x$-direction without loss of generality. Now denoting the wave vector in the fluid rest frame by $(\Omega_\star,\mathbf{k}_\star)$ and the wave vector in the boosted frame by $(\Omega,\mathbf{k})$, we obtain the new dispersion relations by taking $\Omega\rightarrow\Omega_\star$ and $k\rightarrow k_\star$ in \eqs\eqref{rfpm} and \eqref{om4rest}, and then using $\Omega_\star=\gamma\Omega-\gamma V k_x$ and $k^{2}_\star=\gamma^2(\Omega V-k_{x})^2-k_T^2$ with $k_T\equiv \sqrt{k_y^2+k_z^2}$.
 We find for the new dispersion relation $\Omega(k)$ of the longitudinal spin modes in the boosted frame from the first term in \eq\eqref{rfpm}
\begin{equation}
   \gamma^2 [\mathfrak{c}_0-(\mathfrak{c}_1+\mathfrak{c}_2)V^2]\Omega^2+[-i\gamma-2\mathfrak{c}_0\gamma^2 V k^x+\gamma^2(\mathfrak{c}_1+\mathfrak{c}_2)2Vk^x]\Omega+i \gamma V k^x-(\mathfrak{c}_1+\mathfrak{c}_2)(\gamma^2k_x^{2}-k_T^2)+\mathfrak{c}_0 \gamma^2 V^2 k_x^{2}=0\; , \label{oopm}
\end{equation}
and the same equation with $\mathfrak{c}\rightarrow\mathfrak{d}$ from the second term.
For the transverse spin modes we obtain from \eq\eqref{om4rest}
\begin{align}
& \left[\mathfrak{c}_0\mathfrak{d}_0+\gamma^4V^4\mathfrak{c}_2\mathfrak{d}_2+-\mathfrak{c}_{(0}\mathfrak{d}_{2)}\gamma^4 V^2\right]\Omega^{4} +\left[-i\mathfrak{c}_0-i\mathfrak{d}_0-4\gamma^4\mathfrak{c}_0\mathfrak{d}_0 V k^x+\mathfrak{c}_{(0}\mathfrak{d}_{2)}4\gamma^4 V^3k^x+ (i\mathfrak{d}_2+i\mathfrak{c}_2)\gamma^3 V^2\right.\n\\
&\left. -4\mathfrak{c}_2\mathfrak{d}_2 \gamma^4 V^3 k_x \right]\Omega^{3}+\left[\mathfrak{c}_0\mathfrak{d}_0\gamma^46 V^2 k_x^2-3(-i\mathfrak{c}_0-i\mathfrak{d}_0)\gamma^3Vk_x-1-\mathfrak{c}_{(0} \mathfrak{d}_{2)} (\gamma^4V^4k_x^2+4\gamma^4V^2k_x^2+\gamma^4k_x^2-\gamma^2k_T^2)\right.\n\\
&\left.+(i\mathfrak{d}_2+i\mathfrak{c}_2)(-\gamma^3 V^3k_x-2\gamma^3Vk_x)+\mathfrak{d}_2\mathfrak{c}_2(6\gamma^4 V^2 k_x^2-2\gamma^2 V^2k_T^2)-\mathfrak{c}_3\mathfrak{d}_3\gamma^2V^2\right]\Omega^{2}+\left[-4\mathfrak{c}_0\mathfrak{d}_0\gamma^4 V^2 k_x^3\right.\n\\
 &\left. -(i\mathfrak{c}_0+i\mathfrak{d}_0) 3\gamma^3 V^2 k_x^2+2\gamma^2 V k_x-\mathfrak{c}_{(0} \mathfrak{d}_{2)} (-2\gamma^4 V^3 k_x^3-2\gamma^4 k_x^2 V+2\gamma^2 V k_x k_T^2)+\mathfrak{c}_2\mathfrak{d}_2(-4\gamma^4 V k_x^3+4\gamma^2Vk_x k_T^2)\right.\n\\
 &\left. +2\mathfrak{c}_3\mathfrak{d}_3\gamma^2V k^x \right]\Omega+\mathfrak{c}_0\mathfrak{d}_0\gamma^4 V^4 k_x^4-(-i\mathfrak{c}_0-i\mathfrak{d}_0)\gamma^3 V^3 k_x^3-\gamma^2V^2 k_x^2-\mathfrak{c}_{(0} \mathfrak{d}_{2)} (\gamma^2k_x^2-k_T^2)\gamma^2 V^2 k_x^2\n\\
 &-(i\mathfrak{d}_2+i\mathfrak{c}_2)(\gamma^2k_x^2-k_T^2)\gamma V k_x+\mathfrak{c}_2 \mathfrak{d}_2 (\gamma^4 k_x^4-2\gamma^2k_x^2 k_T^2+k_T^4)-\mathfrak{c}_3\mathfrak{d}_3 (\gamma^2 k_x^2-k_T^2) =0 \; . \label{transmove} 
\end{align}
In the following, we will discuss only the longitudinal spin modes for the sake of illustration of the arguments given in the main text. Although the solution of \eq\eqref{transmove} for the transverse spin modes can also be straightforwardly obtained, it is not very enlightening due to its length.

As an example, consider \eq\eqref{oopm} for small wave number. The nonhydrodynamic modes for $k\rightarrow 0$ are given by
\begin{equation}
   \gamma^2 [\mathfrak{c}_0-(\mathfrak{c}_1+\mathfrak{c}_2)V^2]\Omega^2-i\gamma\Omega=0\; ,
\end{equation}
with the result
\begin{equation}
 \Omega_{\parallel1,\text{nh}}= \frac{i}{\gamma [\mathfrak{c}_0-(\mathfrak{c}_1+\mathfrak{c}_2)V^2]}\; ,  \qquad \qquad (k\rightarrow 0) \; . 
\end{equation}
We see that in this limit the nonhydrodynamic modes are stable given that both the stability conditions \eqref{cond1} and the causality conditions \eqref{cond2} in the rest frame are fulfilled.

We now show that the conditions \eqref{cond1} and \eqref{cond2} are sufficient to ensure causality of the longitudinal modes in the boosted frame for any wave number. Assume that \eqs\eqref{cond1} and \eqref{cond2} hold and consider the causality condition for the dispersion relation for the longitudinal modes in a general frame, given by \eq\eqref{oopm}. Since $\Omega\sim k$ in the limit of large $k$, \eq\eqref{oopm} becomes for $k\rightarrow \infty$
\begin{equation}
   \gamma^2 [\mathfrak{c}_0-(\mathfrak{c}_1+\mathfrak{c}_2)V^2]\Omega^2-2\gamma^2V[\mathfrak{c}_0 -(\mathfrak{c}_1+\mathfrak{c}_2)]k^x\Omega-(\mathfrak{c}_1+\mathfrak{c}_2)(\gamma^2k_x^{2}-k_T^2)+\mathfrak{c}_0 \gamma^2 V^2 k_x^{2}=0\; ,
\end{equation}
which has the solution
\begin{align}
\Omega_{\parallel1}
     &=\frac{\mathfrak{c}_0-(\mathfrak{c}_1+\mathfrak{c}_2)}{\mathfrak{c}_0-(\mathfrak{c}_1+\mathfrak{c}_2)V^2} V \cos \theta k\n\\
        &\pm \frac{1}{\mathfrak{c}_0-(\mathfrak{c}_1+\mathfrak{c}_2)V^2}\sqrt{(-2V^2+1+V^4) \mathfrak{c}_0(\mathfrak{c}_1+\mathfrak{c}_2)\cos^2\theta-\frac{1}{\gamma^2}[\mathfrak{c}_0-(\mathfrak{c}_1+\mathfrak{c}_2)V^2](\mathfrak{c}_1+\mathfrak{c}_2)\sin^2\theta}\, k\; .
\end{align}
where we defined $k_x\equiv k\cos\theta$ and $k_T\equiv k\sin\theta$.
The term under the square root is maximal for $\theta=0$, therefore we have
\begin{align}
    v_g&\leq \frac{1}{\mathfrak{c}_0-(\mathfrak{c}_1+\mathfrak{c}_2)V^2} \left\{[\mathfrak{c}_0-(\mathfrak{c}_1+\mathfrak{c}_2)]V+\frac{1}{\gamma^2}\sqrt{ \mathfrak{c}_0(\mathfrak{c}_1+\mathfrak{c}_2)} \right\}&\leq 1&\n\\
  &\Leftrightarrow \ \mathfrak{c}_0+(\mathfrak{c}_1+\mathfrak{c}_2)V-(1+V)\sqrt{ \mathfrak{c}_0(\mathfrak{c}_1+\mathfrak{c}_2)} &\geq 0&\; . \label{hhhhhhh}
\end{align}
For $V=0$ this reduces to the known condition \eqref{cond1}. On the other hand, for $V=1$ we obtain
\begin{equation}
  \mathfrak{c}_0+(\mathfrak{c}_1+\mathfrak{c}_2)-2\sqrt{ \mathfrak{c}_0(\mathfrak{c}_1+\mathfrak{c}_2)} = (\sqrt{\mathfrak{c}_0}-\sqrt{\mathfrak{c}_1+\mathfrak{c}_2})^2 \geq 0\; .
\end{equation}
Since \eq\eqref{hhhhhhh} is a linear function of $V$, the inequality holds for all $0\leq V \leq1$. Hence, the longitudinal modes are causal in any frame, explicitly demonstrating the statement made in the main text.
  
\end{appendix}

\bibliography{biblio_paper_long}{}

\end{document}